# Spin filtering in transition-metal phthalocyanine molecules from first principles

Li Niu[1], Huan Wang[1], Lina Bai[1], Ximing Rong[2], Xiaojie Liu[1,*], Hua Li[1], Haitao Yin[1,†]

[1]*Key Laboratory for Photonic and Electronic Bandgap Materials of Ministry of Education, School of Physics and Electronic Engineering, Harbin Normal University, Harbin 150025, China*
[2]*Department of Physics and the Center of Theoretical and Computational Physics, The University of Hong Kong, Pokfulam Road, Hong Kong SAR, China*
Corresponding authors. E-mail: *redlxj@163.com, †wlyht@126.com

Using first-principles calculations based on density functional theory and the nonequilibrium Green's function formalism, we studied the spin transport through metal-phthalocyanine (MPc, M=Ni, Fe, Co, Mn, Cr) molecules connected to aurum nanowire electrodes. We found that the MnPc, FePc, and CrPc molecular devices exhibit a perfect spin filtering effect compared to CoPc and NiPc. Moreover, negative differential resistance appears in FePc molecular devices. The transmission coefficients at different bias voltages were further presented to understand this phenomenon. These results would be useful in designing devices for future nanotechnology.

## 1. Introduction

To overcome the increasing difficulties and fundamental limitations of conventional Si-based microelectronics, single molecules have been considered as potential building blocks for future nanoelectronic systems. In molecular devices, many interesting physical properties have been predicted theoretically and verified experimentally. Rich phenomena such as spin-valve [1], spin filtering [2], and the Kondo effect [3] can be observed in molecular devices. A considerable amount of effort has been devoted to finding a potential single molecule that can exhibit these interesting properties [4]. For instance, Hao [5] observed spin-filtering and Kondo resonance features by using a MnCu single-molecule magnet bridged between two gold electrodes. On the basis of first-principles calculations, Waldron [6] predicted a magnetoresistance ratio of 27% in the Ni-benzenedithiol-Ni molecular magnetic tunnel junction. These results indicate great prospects for designing a single molecule for high-performance spintronic devices.

For the realization of organic spintronic devices, benzenedithiol, porphyrin, and phthalocyanine molecules with perfectly symmetrical geometric structures have been considered as strong candidate materials. Especially, transition metal phthalocyanine (MPc), which has a transition metal center and four coordinated pyrrole aromatic rings joined by nitrogen, has gained extensive attention because of its remarkably high thermal stability and the possibility to tune its electronic structure and transport properties by changing the central metallic cation. Some applications of the molecules, such as gas sensors [7], organic light-emitting devices [8], cancer therapy [9], and switches [10,11], have been reported. The transport properties of many devices based on transition metal phthalocyanine have been studied theoretically and experimentally. One question arises: is it possible to fabricate phthalocyanine molecule-based spintronic devices that exhibit spin-dependent transport properties? To address this issue, a substantial amount of effort has been devoted to studying phthalocyanine molecule-based devices. For example, Stefan Schmaus[12] found that the magnetoresistance across a single hydrogen phthalocyanine molecule in contact with a scanning tunneling microscope (STM) ferromagnetic tip is 60%. Hsu [13] investigated spin-polarized tunneling spectra of a single manganese phthalocyanine (MnPc) molecule adsorbed on a Co nanoisland in a spin-polarized STM measurement and demonstrated that the tunnel magnetoresistance (TMR) effect can be efficiently turned by the tip-molecule distance. Recently, Cui [14] predicted that a large spin polarization and negative differential resistance (NDR) can coexist in an iron-phthalocyanine (FePc) molecule sandwiched between two zigzag graphene nanoribbon electrodes.

This work focuses on studying spin-polarized transport through transition metal phthalocyanine (MPc, M=Mn, Fe, Co, Ni, Cr) sandwiched between two semi-infinite Au(001) nanowires based on the nonequilibrium Green's function (NEGF) formalism and density functional theory (DFT). We paid more attention to the effects caused by changing the central metallic cation. We found that the MnPc, FePc, and CrPc molecules exhibit perfect spin filtering effects. The transmission spectra were also presented in order to understand the physics of the electron transport.

## 2. Simulation model and calculation method

The molecular device of Au(001)-MPc-Au(001) is illustrated in Figure 1, where a MPc molecule is sandwiched by two semi-infinite aurum (001) nanowire leads. The entire model system consists of three parts: left electrode, right electrode, and central region. The left and right electrodes are identical and composed of aurum atoms that extend to infinity in both the left and right directions. In this study, the electrodes are quasi-one-dimensional aurum wires, composed of periodic units repeated along the face-centered cubic (001) direction. The repeating unit consists of two layers that alternate between containing four and five aurum atoms. It is well-known that the interaction of leads and the MPc molecule is very weak because of the absence of a direct chemical bond. Here, we slightly modified the geometry of the system, connecting the MPc molecule and leads through two sulfur-bridges. The choice of sulfur functionalization is justified by the well-known ability of thiol groups to anchor molecules to gold electrodes [15]. The sulfur atoms in the molecule are coordinated to the four-aurum atom layer. The structures of the Au nanowire and every single MPc molecule are first optimized independently, and then the whole transport system is optimized further with VASP [16], employing the local density approximation (LDA) for the exchange-correlation functional. We fix the position of the left aurum electrode and relax the MPc molecule and the right aurum electrode. All the structures are relaxed until a force tolerance of 0.05 eV/Å is reached. We find that the total energy is a minimum when the distance between the electrode and sulfur atom reaches 2.2617 Å along the $z$-direction (transport direction).

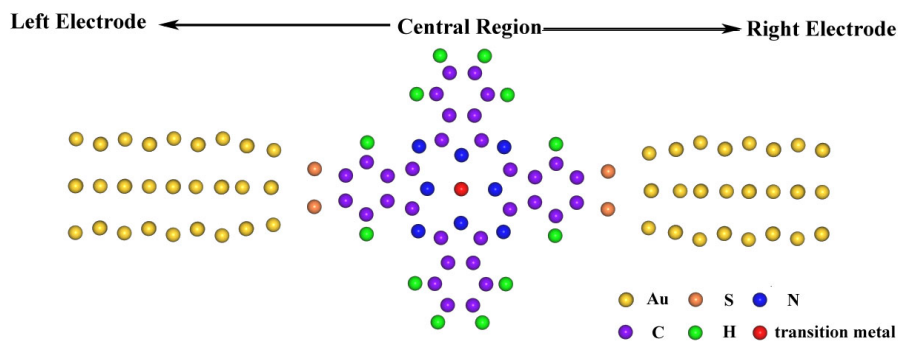

**Fig. 1**    Schematic plot of the Au(001)-MPc-Au(001) device.

The electronic transport properties in the above structure are investigated with Nanodcal [17], which combines DFT with NEGF technique. The DFT method provides the system electronic structure, and the NEGF includes information of the nonequilibrium quantum statistics. Here, the double-zeta polarized (DZP) basis set is used for all the elements and atomic core, which was defined by the standard nonlocal norm-conserving pseudopotential [18]. The exchange-correlation is considered with LDA. A sufficient amount of buffer layers for the Au atomic electrode were included in the central region so the boundary conditions of the scattering region could be determined through the corresponding infinite left and right electrodes. In our calculation, the DFT-NEGF self-consistency is carried out until the numerical tolerance of the Hamiltonian is less than $5\times 10^{-5}$ eV. In this structure, the two aurum (001) atomic electrodes extend to $z=\pm\infty$, where bias voltages are applied and the charge current is collected.

For the two-probe system, the spin-dependent current can be obtained using the Landauer–Büttiker formula [17]

$$I_\sigma(V) = \frac{e}{h}\int_{\mu_L}^{\mu_R} dE T_\sigma(E,V)\left[f_L(E,\mu_L) - f_R(E,\mu_R)\right], \tag{1}$$

where $\sigma \equiv \uparrow, \downarrow$ is the spin index, and $\mu_L(\mu_R)$ and $f_L(f_R)$ are the electrochemical potentials and the Fermi distribution function of the left (L) and right (R) leads, respectively. The relation of the bias voltage $V$ and electrochemical potential satisfies $\mu_L - \mu_R = eV$. $T_\sigma(E,V)$ is the spin-resolved transmission coefficient, defined as

$$T_\sigma \equiv Tr\left[\text{Im}(\Sigma_L^r) G^r \text{Im}(\Sigma_R^r) G^a\right], \tag{2}$$

where $\Sigma_{L/R}^r$ are the retarded self-energy of the left or right leads, which reflect the coupling between the leads and the central scattering region; and $G^{r/a}$ are the retarded (advanced) Green's function matrices in spin and orbital space. The total charge current flowing through the device is $I \equiv I_\uparrow + I_\downarrow$.

The spin filter efficiency (SFE) [19,20] at the voltage is defined as

$$\text{SFE} = \frac{|I_\uparrow(V) - I_\downarrow(V)|}{I_\uparrow(V) + I_\downarrow(V)}, \tag{3}$$

which represents the excess current of one spin type over the other as a percentage of the total current.

## 3. Results and discussion

After obtaining the most stable configurations, we can calculate the current–voltage (I–V) curves for different spin electrons according to the Landauer–Büttiker formula. Figures 2(a)–(e) present the current-flowing phthalocyanine molecule with different central cations between the two electrodes. The SFE at different bias voltages is also presented in Figure 3. Note that the SFE is calculated using the equation $\text{SFE}=\left(\left|T_\uparrow(0)-T_\downarrow(0)\right|\right)/\left(T_\uparrow(0)+T_\downarrow(0)\right)$ in the equilibrium state. The distinct features of the transport properties in this structure are as follows: (i) The currents show the same tendency even though the central metallic cations are different. The currents increase when the bias voltages increase from zero. (ii) The SFE reaches about 80% for CrPc and even reaches 100% for MnPc and FePc. This implies that for MnPc and FePc a perfect spin filter effect is presented, e.g. only one spin state electron flowing through the device can be allowed, another spin state electron is forbidden. (iii) The SFE of CoPc is very small compared to that of CrPc, MnPc, and FePc. Especially, for NiPc, the SFE is zero. (iv) For FePc, the currents decrease for increasing bias voltage in the range of (0.1, 0.2)V, which implies that the well-known phenomenon of NDR appears.

**Fig. 2** Current-flowing MPc molecular devices versus bias voltage with different central cations: **(a)** MnPc; **(b)** CrPc; **(c)** FePc; **(d)** CoPc; **(e)** NiPc. Transmission coefficient versus energy at zero bias for **(f)** MnPc; **(g)** CrPc; **(h)** FePc; **(i)** CoPc; **(j)** NiPc.

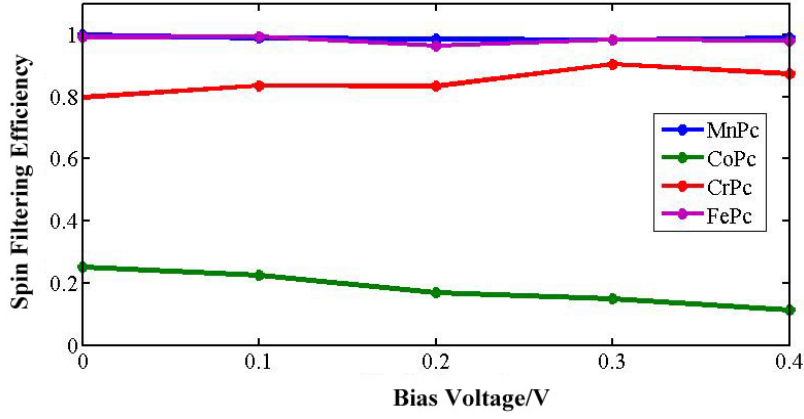

**Fig. 3** Spin filtering efficiency at different voltages.

The voltage dependence of the spin current can be understood from the behavior of the transmission coefficient. Figs 2(f)–(j) plot the transmission coefficient $T_\sigma$ as function of energy $E$ near the Fermi energy at zero bias. Note that the $T_\downarrow$ curves of MnPc and FePc are dominated by sharp peaks and $T_\uparrow$ is about zero at the Fermi energy. A fully polarized peak of the CrPc transmission coefficient emerges at about 0.24 eV. For NiPc, the transmission coefficient of different spin states is entirely equal and no peak emerges near the Fermi energy. The characteristics of the spin-dependent transmission coefficient are the immediate cause of the SFE at nonequilibrium. When the positive bias voltage is applied to the right electrode (we set $\mu_L = 0$ in our calculation), the current is contributed from the energy integral interval (–eV, 0), according to Eq. (1). Moreover, the entire transmission spectrum shifts to low energies but roughly maintains its shape. As the bias voltage V is increased, the transport window (–eV, 0) becomes wider. Hence, the peak labeled by "A" in the spin-down channel enters the bias voltage window to cause an increase in the curve of $I_\downarrow$, as shown in Figs 2(f)–(h). As a result, a perfect SFE emerges in the MnPc, FePc, and CrPc devices. For CoPc, there exists a spin-polarized peak at –0.4 eV, but the peak is far from the Fermi energy and only the right tail of the peak can contribute to the current under bias voltage, which leads to a very small current and SFE. For NiPc, the transmission probabilities for the two spin components are identical and no peak emerges near the Fermi energy, as shown in Fig. 2(j), which straightly leads to a very small and non-spin-polarized current. At a larger bias voltage, the left tail of the peak near 0.5 eV in the transmission spectrum may enter the bias voltage window. Accordingly, the current will increase rapidly.

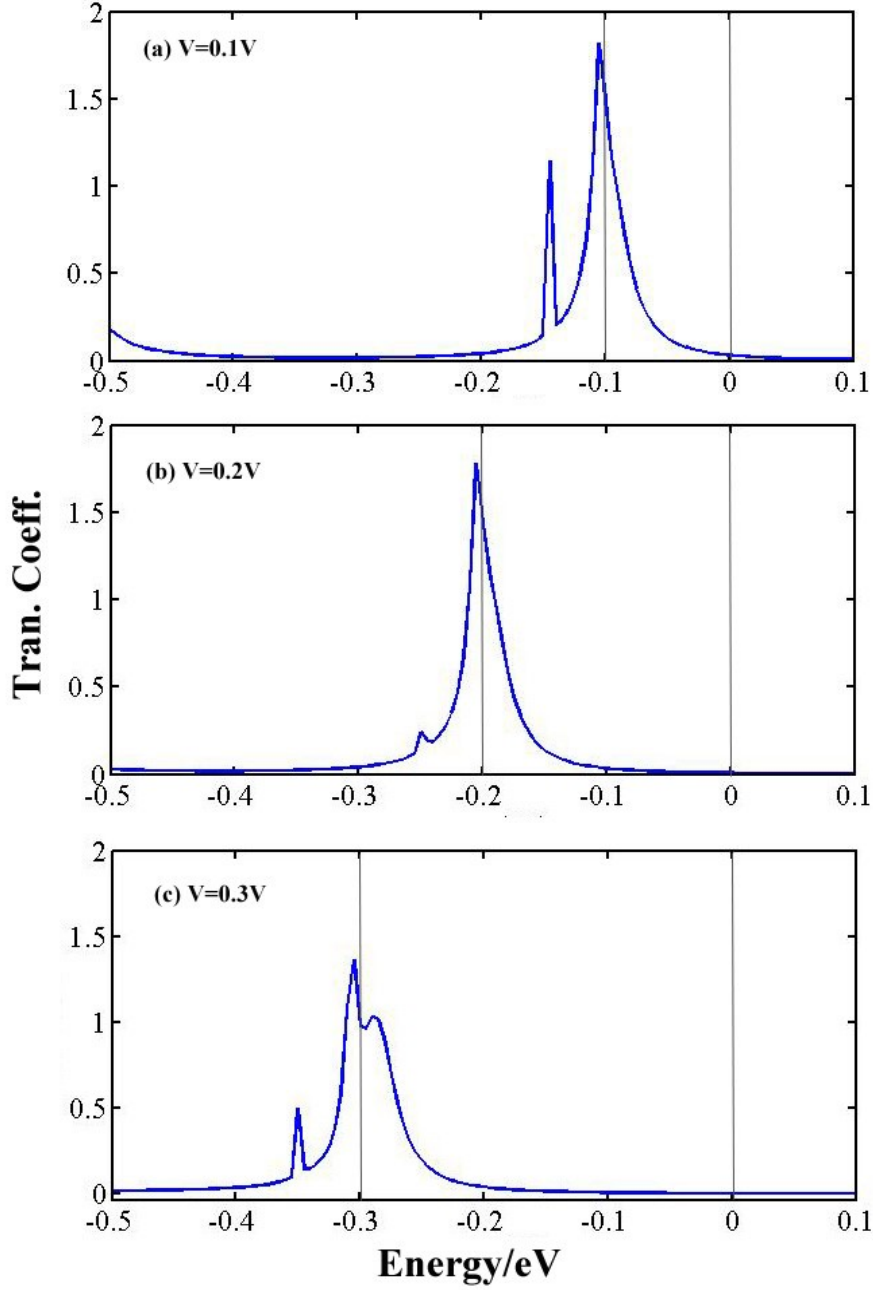

**Fig. 4** Transmission coefficient versus energy at different bias voltages.

Finally, to understand the phenomenon of NDR in the *I–V* curves of FePc, we plot the transmission coefficient as a function of energy at different bias voltages, as shown in Fig. 4. The currents are determined by the integration interval (–*eV*, 0) and the transmission coefficient $T_\sigma(E,V)$, and the latter is also a function of the bias voltage and energy. Therefore, the situation becomes more complicated. In Fig 4, the bias voltage window is denoted by the area between the two vertical lines. The current can be determined by the integral contribution of the transmission coefficient over the bias voltage window (–*eV*, 0). We find that the integral value of the transmission coefficient at 0.2 V is slightly smaller than that of transmission coefficient at 0.1 and 0.3 V. Therefore, the current value decreases when the value of the bias voltage increases from 0.1 to 0.2 V and the NDR appears in this voltage region.

4. **Conclusions**

In summary, we have studied the spin transport through metal-phthalocyanine (MPc, M=Ni, Fe, Co, Mn, Cr) molecules connected to aurum nanowire electrodes within the DFT-NEGF method. The transmission coefficient at equilibrium and nonequilibrium as well as the current–voltage curves were calculated. We found that MnPc, FePc, and CrPc molecular devices exhibit perfect spin filtering effects. By contrast, the spin-filtering efficiency of CoPc is very small and even that of NiPc is zero. The transmission coefficients at different bias voltages were further presented to understand the negative differential resistance appearing in the FePc's current–voltage curves. These results helped clarify the transport properties of molecular devices and would be useful in designing devices for future nanotechnology.

**Acknowledgements** This work was financially supported by the opening project of Key Laboratory for Photonic and Electronic Bandgap Materials of Ministry of Education and the National Natural Science Foundation of China (Grant No. 11504072).